# A First-Principles Study of CdSe Nanoclusters Capped by Thiol Ligands

*Shanshan Wu[1,2,]\*, Michael McGuigan[2], Amanda L. Tiano[3], Stanislaus S. Wong[3,4], and James G. Glimm[1,2]*

1   Department of Applied Mathematics and Statistics, Stony Brook University, Stony Brook, NY, 11794-3600;

2   Computational Science Center, Brookhaven National Laboratory, Building 463, Upton, NY, 11973-5000;

3   Department of Chemistry, Stony Brook University, Stony Brook, NY, 11794-3400;

4   Condensed Matter Physics and Materials Sciences Department, Brookhaven National Laboratory, Building 480, Upton, NY, 11973;

\*   Address: Department of Applied Mathematics & Statistics, Stony Brook University, Stony Brook, NY, 11794-3600; Email: sunshine.wu0602@gmail.com, Phone: +1-631-413-4947, Fax: +1-631-632-8490.

**Abstract:** A first-principles study of small $Cd_nSe_n$ quantum dots (QDs) ('n' = 6, 13, and 33) has been performed for the study of QD-sensitized solar cells. We assessed the effects of the passivating thiol-radical ligands on the optimized structure, the energy gap, and on the absorption spectrum. The simplest thiol, methanethiol, and four other thiol type ligands, namely — cysteine (Cys), mercaptopropionic acid (MPA), and their reduced-chain analogues, were investigated. We have come to the following conclusions. (a) Thiol-radical ligands possessed greater effects on the structure and electronic properties of the CdSe QDs than thiol ligands alone. (b) The sulfur 3p orbitals were localized as the midgap states for the thiol-radical-ligated complex, which altered the absorption spectrum of bare $Cd_6Se_6$ by inducing a new lower energy absorption peak at 2.37 eV. (c) The thiol-radical-ligated complex was also found to be sensitive to the position and number of ligands. (d) Both the amine group on Cys and the carboxyl group on Cys and MPA showed a strong tendency to bond with the neighboring Cd atom, especially when the length of the ligand was reduced. This formation of Cd-N and Cd-O bonds resulted in smaller HOMO-LUMO gaps and a stronger binding between the ligands and the surface atoms of CdSe nanoclusters.

## 1. Introduction

Quantum-dot-sensitized solar cells[1] (QDSCs) have been investigated intensively.[2] The quantum dots (QDs) possess the ability to tune their absorption spectrums by size-control[3-6] and surface manipulation.[7-9] As a result of their properties, QDs have the potential to produce a composite solar cell which fully covers the solar spectrum and to obtain a conversion efficiency exceeding the Shockley–Queisser limit of 32.9%.[10,11] The CdSe–TiO$_2$ composite QDs are commonly used materials for QDSCs.[2-9, 12-15] The effect of surface passivation of CdSe QDs on the electronic and optical properties of the device is one of the central issues for the study of the CdSe-TiO$_2$ QDSCs. Many theoretical[16-22] and experimental[7-9,15] studies have been performed. Kuznetsov *et al.*[22] provided a detailed summary of previous theoretical studies of the structural and electronic properties of small bare and capped CdSe QDs. In a newly published experimental study from Nevins *et al.*,[9] the amine group of cysteine (Cys) was reported to play a crucial role in producing the less than 2nm ultra stable CdSe QD, in contrast with the ligand of mercaptopropionic acid (MPA), which only produced a regular larger size CdSe QD. A narrower peak and a higher absorption intensity were also observed for the Cys-capped CdSe QD, while the effects of the amine group on the absorption properties were not fully understood.

In this paper, we report a comprehensive study of the CdSe nanoclusters capped by the thiol family ligands (methanethiol (MT), Cys and MPA) using first-principles density functional theory (DFT) and time-dependent DFT (TDDFT), which evinced some interesting results. First, the ligands in the thiol-radical form bound 62% more strongly to the CdSe nanoclusters than did in the thiol form. Secondly, we found that the thiol-radical ligands also had a greater effect on the electronic and optical properties of QDs. The photo-physical properties of CdSe nanoclusters were altered by introducing new lower energy absorption peaks when passivated by the thiol-radical ligands. When passivated by the thiol ligands, we observed only a small blue shift of the absorption spectrum by 0.2 eV. We also noted that altering the position and number of thiol-radical MTs attached to the $Cd_6Se_6$ resulted in significant changes to the structure and corresponding electronic properties. Lastly, our work revealed that both amine and carboxyl groups on the thiol-radical ligands exhibited a strong tendency to bond with the nearby Cd atom, especially when the length of ligand was reduced. The formation of Cd-N and Cd-O bonds resulted in smaller energy gaps and a stronger binding between the ligands and the surface atoms of CdSe nanoclusters.

The paper is organized as follows: In Section 2, we introduce the simulation model and methodology. In Section 3, we present and analyze the simulation results. We summarize our findings in Section 4.

## 2. Simulation Details

*2.1 CdSe Cluster and Ligand Models*

Four sizes of $Cd_nSe_n$ QDs with 'n' = 6, 12, 13, and 33 were cut directly from the CdSe würtzite bulk crystal.[23] The würtzite structure of CdSe QDs has been commonly used in experiments[2,4,9] and has also been intensively studied from a theoretical point of view.[16,18-22]

This paper was focused on a family of thiol ligands. This family is commonly used as the ligands of CdSe QDs. Specifically, MT ($CH_3SH$), which is the shortest thiol ligand, was chosen to study the differences amongst the thiol, thiol-radical, and thiolate species. We also used the MT thiol-radical ligand to study the effects of changing the position and number of ligands attached to the CdSe nanoclusters. We compared the CdSe nanoclusters capped by the MPA and Cys thiol-radicals, respectively. Both MPA and Cys possess a common structural backbone consisting of a thiol group and terminating with a carboxyl group (e.g. HS-R-COOH). For MPA, the 'R' functional group is a pure alkane chain, while the 'R' function group of Cys maintains an extra amine group attached to the alkane chain. We also carried out another complementary, comparative study of the effects of the actual lengths of the ligands themselves by reducing the length of the 'R' chains of MPA and Cys.

According to prior experimental results,[9,13,15] both the thiol-radical and thiolate species possess a high affinity for the Cd atom. In this paper, we restricted our research to the gas phase complexes. Though the surrounding organic solvents can influence the binding strength of ligands on CdSe surfaces,[16,24] our main focus has been to study the effects of the thiol-radical ligands, bound to the CdSe QDs, on structural relaxation and electronic properties.

*2.2 Methodology*

Our calculations used the NWCHEM 6.0 program[25] with basis sets, as well as LANL2DZ[26] and 6-31G*[27,28] potentials for CdSe and ligands respectively. The LANL2DZ effective core potential[26] was used for the CdSe core atoms. This choice of basis sets has proved to be a pragmatic but sufficient and necessary balance between accuracy and computational cost.[24] The B3LYP[29] exchange and correlation (XC) functional was applied for DFT geometry optimization and the TDDFT calculation. To reduce the energy state degeneracy, the symmetry was suppressed during the simulation. This appropriateness of our choice has been verified in previous work.[20-22] We also confirmed the methodology by carrying out a study of the bare $Cd_nSe_n$ nanoclusters ('n' = 6, 12, 13, and 33) (Supplementary material, Section 1) by comparison with previous work of others.

*2.3 Ligand Binding Energies*

The binding energy (BE), is defined as the averaged binding energy per ligand molecule:

$$BE = \frac{1}{m}\left(E_{Cd_nSe_nL_m} - E_{Cd_nSe_n} - mE_L\right),$$

where 'n' = 6, 13, and 33; 'm' is the number of ligands; and $E_{Cd_nSe_nL_m}$, $E_{Cd_nSe_n}$ and $E_L$ are associated with the total energies of the ligand-bound system, the bare quantum dot and the single ligand, respectively. According to the methodology work from Albert, et al.,[24] LANL2DZ/6-31G* represents the minimal basis set needed to obtain accurate QD-ligand interactions without further correction for the basis set superposition error.

**3. Simulation Results and Discussion**

To investigate the structural and optical properties of thiol family ligated CdSe nanoclusters, we started by comparing these properties for the thiol, thiol-radical, and thiolate ligands bound to the $Cd_6Se_6$ nanoclusters (Section 3.1). Then, we studied the effects of variation in the number and position of ligands bonded to the $Cd_nSe_n$ nanoclusters ('n' = 6 and 13) (Section 3.2). Finally, we compared the $Cd_6Se_6$ nanoclusters capped by MPA, Cys, and their reduced chain analogues (Section 3.3).

*3.1. Thiol, Thiol-Radical and Thiolate*

We considered (1) an optimized $Cd_6Se_6$ + $SHCH_3$ complex structure, (2) removal of the hydrogen atom attached to the sulfur, which resulted in a thiol-radical ligated complex, and (3) deprotonation of the thiol ligand so that the whole system became a thiolate-ligated complex. First, we studied the three complexes without further geometry relaxation. The BEs of the Cd-S bond for these three complexes were -12.9, -13.4, and -66.7 kcal/mol, respectively (Table 1). As shown in Figure 1L, when the thiol was dehydrogenated into the thiol-radical, one spin of the sulfur $3p$ orbitals was emptied, and the empty orbital localized as a midgap state close to the HOMO state (Table 2). This midgap state resulted in the presence of new lower energy absorption peaks around 1 eV (red line in Figure 1L). When the thiol was deprotonated into thiolate, the sulfur $3p$ orbital was occupied by the extra electron and localized as the midgap states between the HOMO and LUMO state (Table 2). These midgap states resulted in the presence of new lower energy absorption peaks around 2 eV (Figure 1L).

Then, we relaxed the thiol-radical and thiolate-ligated complexes, respectively. As shown in Table 1, after relaxation, we observed a larger difference among the Cd-S BEs for the three composed systems. As compared with the original thiol-ligated complex, the Cd-S bond lengths of the relaxed thiol-radical ligated complex and the thiolate-ligated one were shortened by 7% and 11%, respectively. The HOMO-LUMO gaps of the two complexes were also increased as a result of the structural relaxation. For the thiol-radical and thiolate-ligated complexes, the sulfur $3p$ orbitals were still localized around the frontier orbitals and altered the absorption spectrum of the bare $Cd_6Se_6$ nanocluster by introducing a new lower energy absorption peak (Figure 1R). By contrast, for the thiol-ligated complex, the sulfur orbitals were localized deep inside the valence and conduction bands (Supplementary Material, Figure S2), while the first and second absorption peaks were blue shifted from those of the bare nanocluster by 0.2 eV, respectively (Supplementary Material, Figure S3L).

From these simulations, we concluded that thiol-radical and thiolate species were bonded more strongly than thiol to the CdSe surface (Table 1). Their sulfur 3p orbitals generated midgap states (Table 2), which altered the absorption spectrum of bare $Cd_6Se_6$ by inducing a new lower energy absorption peak (Figure 1R).

Table 1. BE (kcal/mol) and bond length (Å) of Cd-S bond (d(Cd-S)), and HOMO-LUMO gap (eV) of $Cd_6Se_6$ + $SHCH_3$, $SCH_3$ and $SCH_3^-$. The geometry optimization was performed by using the B3LYP functional with the LANL2DZ/6-31G* (CdSe/Ligand) basis sets.

|  |  | Bare | $SHCH_3$ (Thiol) | $SCH_3$ (Thiol-Radical) | $SCH_3^-$ (Thiolate) |
|---|---|---|---|---|---|
| **Structures Derived From Relaxed $Cd_6Se_6$ + $SHCH_3$** | BE | -- | -12.91 | -13.38 | -66.71 |
|  | d(Cd-S) | -- | 2.817 | 2.817 | 2.817 |
|  | HOMO-LUMO Gap | 3.06 | 3.23 | 3.26(α)/0.84(β) | 1.25 |
| **Optimized Complexes** | BE | -- | -12.91 | -20.91 | -76.94 |
|  | d(Cd-S) | -- | 2.817 | 2.604 | 2.516 |
|  | HOMO-LUMO Gap | 3.14 | 3.23 | 2.91(α)/2.58(β) | 2.51 |

Table 2. Density of states (DOS) plots: Left (L), optimized $Cd_6Se_6$ + $SHCH_3$ system; Middle (M), optimized thiol-radical-ligated system; Right (R), optimized thiolate-ligated complex The B3LYP / (LANL2DZ/6-31G*) method is used for the DFT calculation. A Gaussian broadening of 0.05 eV has been utilized.

| Structures Derived From Relaxed $Cd_6Se_6$ + $SHCH_3$ | Optimized Complexes |
|---|---|

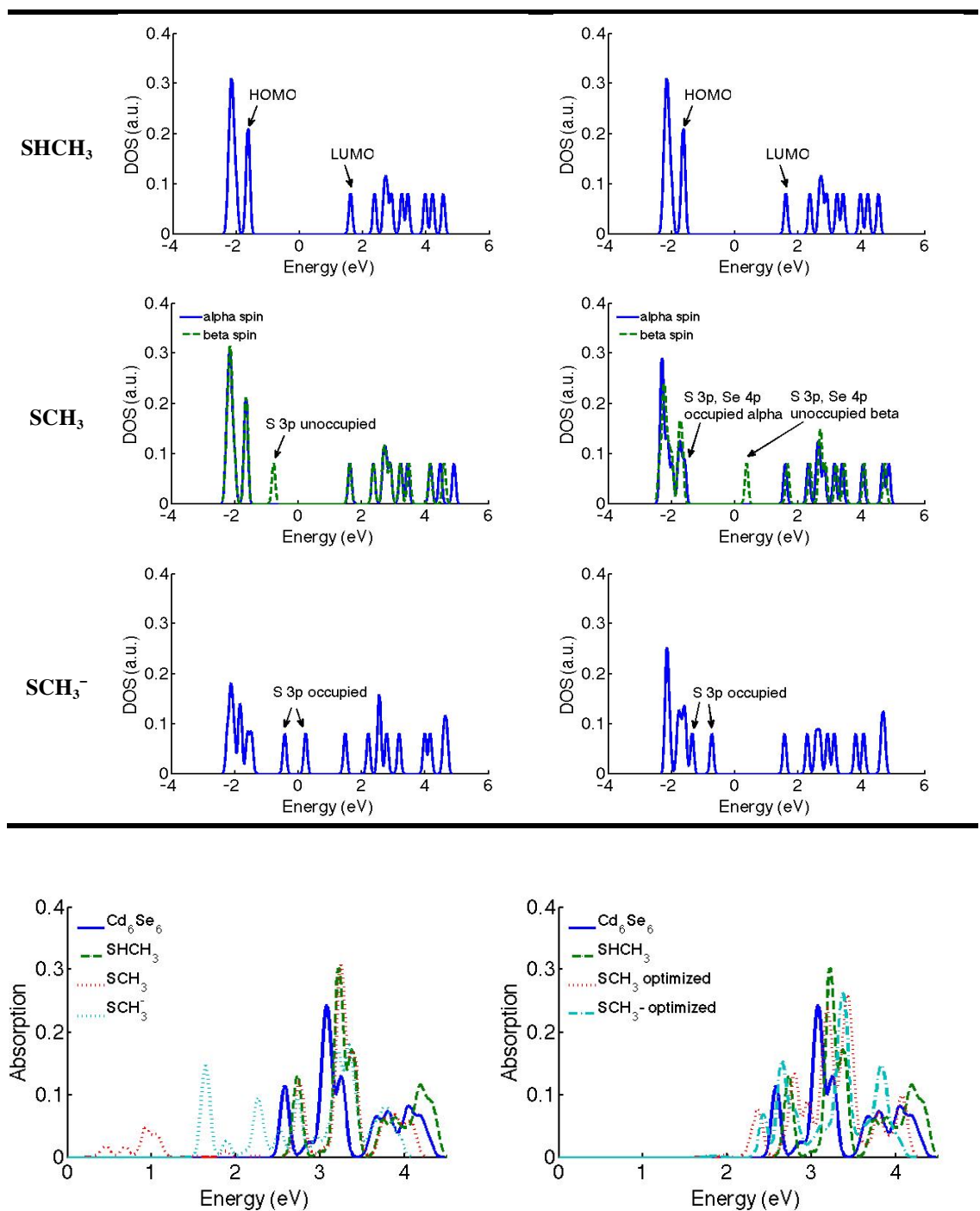

Figure 1. TDDFT spectra plots: Left (L), optimized bare $Cd_6Se_6$, optimized $Cd_6Se_6$ + $SHCH_3$ system, the thiol-radical-ligated and thiolate-ligated complexes without further relaxation; Right (R), with all systems optimized. The B3LYP / (LANL2DZ/6-31G*) method is used for the TDDFT calculation. A Gaussian broadening of 0.05 eV has been utilized.

*3.2. Number and Position of Attached Ligands*

Next, we explored the effects of changing the number and position of the thiol-radical ligands on the electronic and optical properties of QDs. According to the experimental results from Nevins *et al.*,[9] the thiol is deprotonated when attached to the CdSe surface in the solvent. However, due to the presence of positively charged bare CdSe clusters in the solvent, the ligated complexes tend to be neutral.[30,31] Thus, instead of studying the thiolate-ligated CdSe nanocluster, we investigated the thiol-radical-capped charge-neutral CdSe nanocluster.

We saturated the bare $Cd_6Se_6$ with *n* $SCH_3$ ligands ('n' = 1, 2, 3, 6) (Table 3). Since the six Cd atoms possess equivalent positions, we had only one choice when adding either 1 or 6 ligands. However, 2 or 3 ligands can be added in three non-equivalent combinations. All of the QD-ligand composite systems were relaxed to their lowest energy configuration. In each case, the capping ligand always sought out its most stable configuration. The relaxed structures were listed in Table 3. Position 1 (p1) of $Cd_6Se_6$ + 2 $SCH_3$ possessed the lowest energy configuration by a difference of 0.5 eV compared to (Position 2) p2 and 0.3 eV compared to (Position 3) p3, while p2 of $Cd_6Se_6$ + 3 $SCH_3$ maintained the lowest total energy among the three positions by 0.6 eV compared to p1 and 0.1 eV compared to p3 (Table 4).

According to the data from Table 4 for the $Cd_6Se_6$ + 2 $SCH_3$ complexes, we observed a great change of the BE, and HOMO-LUMO gap for the p2 system after the core $Cd_6Se_6$ was unfixed. When the core $Cd_6Se_6$ was fixed, the p2 position of thiol-radical ligands possessed an ultra small BE which is only 1/3 of the BEs of the other two positions. However, when we unfixed the core structure and further relaxed the whole structure, the attached ligands induced a relaxation of the core $Cd_6Se_6$. This change resulted in a more stable structure with a higher BE about -22.19 kcal/mol. For p1 and p2 position, the structures were further stabled by generating a Cd-S-Se bridge.

In Figure 2, we plotted the absorption spectrum for the systems with either 2 or 3 ligands. As we can see, the position of thiol-radical ligands played significant effects on the passivated nanoclusters, while the position and configuration of the thiol ligands was much less sensitive (Supplementary Material Figure S3L). As an illustrative example, in Table S2 and Figure S3L, three thiol ligands were attached to the $Cd_6Se_6$ following the p1 position. The energy gap was opened up by as much as 6% with a corresponding blue shift of the absorption spectrum by 0.2 eV.

When all the Cd atoms of $Cd_6Se_6$ were saturated, the core CdSe relaxed into an open structure with the existence of a Cd-S-Se bridge (Table 3, last row). As shown in the last row of Table 5, the energy gap decreased when the number of $SCH_3$ thiol-radicals attached to the CdSe surface increased. The difference in BEs and bond lengths resulted in the formation of Se-S bonds rather than a change in the number of thiol-radical ligands (Table 5). As shown in Table 6, the number of sulfur 3p orbitals, which localized between the HOMO and LUMO states, was increased when we kept adding the thiol-radical ligands to the surface of CdSe nanoclusters. This observation explained the decrease of energy gaps when increasing the number of ligands.

In the case of $Cd_6Se_6$, all the 6 Cd atoms were equivalent. However, when the size of the QD grows, it may have different facets for the ligand to attach to. For instance, the $Cd_{13}Se_{13}$ cluster maintains three different positions to bond with the ligands (Table 7). Here, we added one $SCH_3$ ligand for each facet to the Cd atom. Each $Cd_{13}Se_{13}$ + $SCH_3$ composite system was then relaxed to its lowest energy configuration. As shown in Table 7, when the ligands were attached in p1 and p2 positions, both the relaxed structures possessed a Cd-S-Se bridge, while the ligand attached to the p3 position was associated with a Cd-S-Cd bridge. The p3 position possesses the lowest energy configuration by 0.03 eV as compared with p1 and 0.1 eV as compared with p2 (Table 4).

Based on our results, we concluded that the electronic properties of the thiol-radical-ligated systems were very sensitive to the precise position of the attached ligand (Figure 2 and Figure 3). The formation of the Cd-S-Se bridge possesses significant effects on the BEs of thiol-radical ligands (Table 5). When the number of attached ligands was increased, the energy gap was further narrowed by the presence of the sulfur $3p$ orbitals (Table 5 and Table 6).

Table 3. Optimized structures of $Cd_6Se_6 + n\ SCH_3$ ('n' = 1, 2, 3, and 6) using the B3LYP functional with the LANL2DZ/6-31G* (CdSe/Ligand) basis sets (Cd: cyan, Se: yellow, H: white, S: orange, C: gray)[i].

|  | Position 1 | Position 2 | Position 3 |
|---|---|---|---|
| $Cd_6Se_6$ + 1SCH$_3$ | 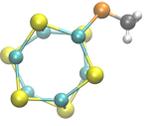 | -- | -- |
| $Cd_6Se_6$ + 2SCH$_3$ | 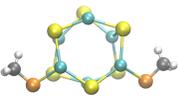 | 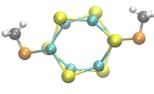 | 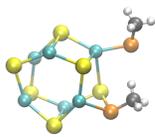 |

---

[i] The images in Table 3, and Table 7 were reduced using VMD.[32]

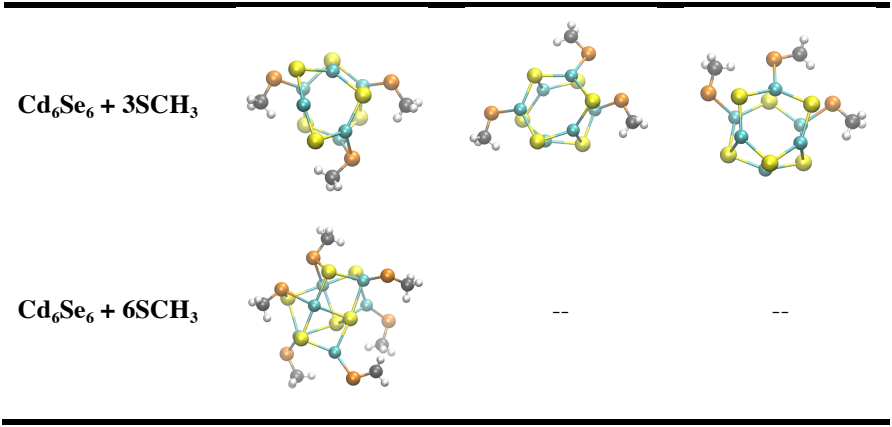

| | Cd$_6$Se$_6$ + 3SCH$_3$ | | |
|---|---|---|---|
| | Cd$_6$Se$_6$ + 6SCH$_3$ | -- | -- |

Table 4. (kcal/mol), bond length (d(Cd-S) and d(Se-S)) (Å) and HOMO-LUMO gap (eV) of optimized structures of Cd$_6$Se$_6$ + $n$SCH$_3$ ('n' = 2, 3) and Cd$_{13}$Se$_{13}$ + 1SCH$_3$ using the B3LYP functional with the LANL2DZ/6-31G* (CdSe/Ligand) basis set.

| | | Position 1 | Position 2 | Position 3 |
|---|---|---|---|---|
| **Cd$_6$Se$_6$ + 2SCH$_3$** (Fixed Core) | BE | -16.70 | -6.88 | -18.20 |
| | *d(Cd-S)* | 2.567 | 2.671 | 2.659 |
| | *d(Se-S)* | -- | 2.541 | -- |
| | HOMO-LUMO Gap | 2.29 | 0.56 | 3.05 |
| **Cd$_6$Se$_6$ + 2SCH$_3$** | BE | -27.79 | -22.19 | -25.79 |
| | *d(Cd-S)* | 2.670 | 2.702 | 2.610 |
| | *d(Se-S)* | 2.521 | 2.341 | 2.593 |
| | HOMO-LUMO Gap | 3.22 | 1.49 | 3.14 |
| **Cd$_6$Se$_6$ + 3SCH$_3$** | BE | -15.63 | -20.27 | -19.11 |
| | *d(Cd-S)* | 2.633 | 2.595 | 2.599 |
| | HOMO-LUMO Gap | 1.91(α)/1.86(β) | 2.25(α)/2.15(β) | 1.84(α)/2.33(β) |
| **Cd$_{13}$Se$_{13}$ + 1SCH$_3$** | BE | -19.87 | -17.17 | -20.57 |
| | *d(Cd-S)* | 2.578 | 2.602 | 2.639 |
| | *d(Se-S)* | -- | -- | -- |
| | HOMO-LUMO Gap | 2.70(α)/2.64(β) | 2.80(α)/2.51(β) | 3.14(α)/1.55(β) |

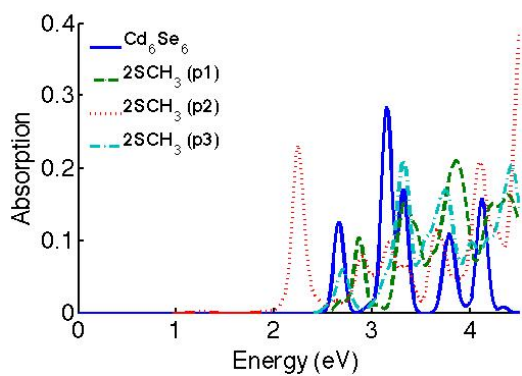
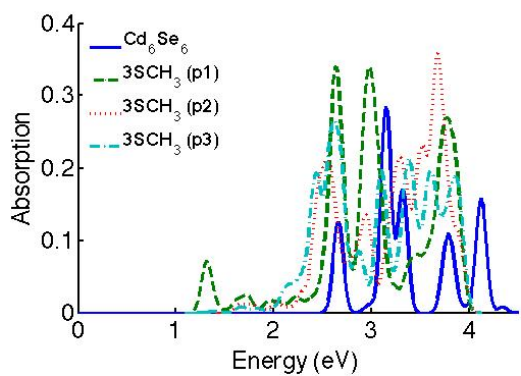

Figure 2. TDDFT spectra plots of different ligand positions of: L, optimized $Cd_6Se_6$ + $2SCH_3$; R, optimized $Cd_6Se_6$ + $3SCH_3$. The B3LYP / (LANL2DZ/6-31G*) method is used for the TDDFT calculation. A Gaussian broadening of 0.05 eV has been utilized.

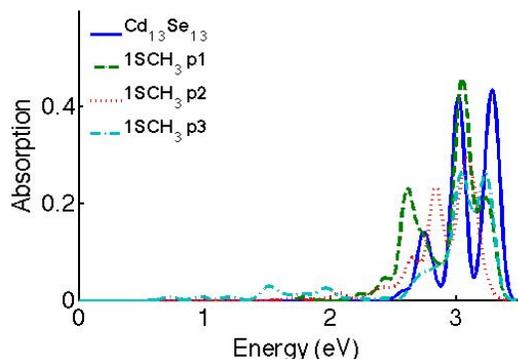

Figure 3. TDDFT spectra plots of optimized $Cd_{13}Se_{13}$ + $1SCH_3$. The B3LYP / (LANL2DZ/6-31G*) method is used for the TDDFT calculation. A Gaussian broadening of 0.05 eV has been utilized.

Table 5. BE (kcal/mol), bond length (d(Cd-S) and d(Se-S)) (Å) and HOMO-LUMO gap (eV) of optimized structures of $Cd_6Se_6$ + $nSCH_3$ ('n' = 1, 2, 3, and 6) using the B3LYP functional with the LANL2DZ/6-31G* (CdSe/Ligand) basis sets.

|  | $1SCH_3$ | $2SCH_3$(p1) | $3SCH_3$(p2) | $6SCH_3$ |
|---|---|---|---|---|
| **BE** | -20.91 | -27.79 | -20.27 | -25.21 |
| *d(Cd-S)* | 2.604 | 2.670 | 2.595 | 2.456 |
| *d(Se-S)* | -- | 2.521 | -- | 2.346 |
| **HOMO-LUMO Gap** | 2.91(α)/2.58(β) | 3.22 | 2.25(α)/2.15(β) | 2.23 |

Table 6. DOS plots of optimized $Cd_6Se_6$ + $n$ $SCH_3$ ('n' = 1, 2, 3, and 6) and the optimized bare $Cd_6Se_6$. The B3LYP / (LANL2DZ/6-31G*) method is used for the DFT calculation. A Gaussian broadening of 0.05 eV has been utilized.

| System | Plots of DOS |
|---|---|
| $Cd_6Se_6$ | 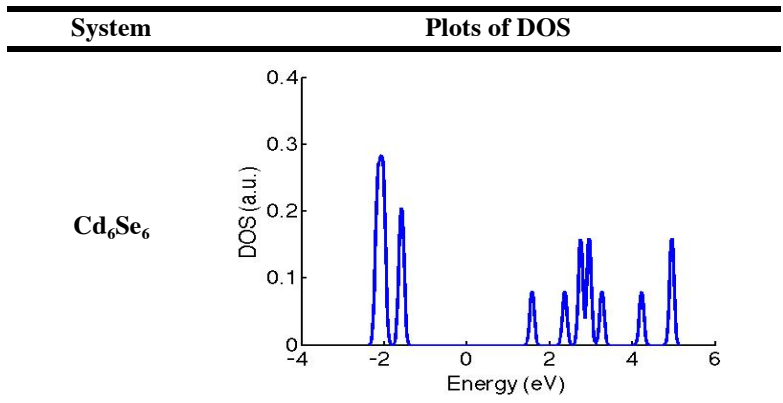 |

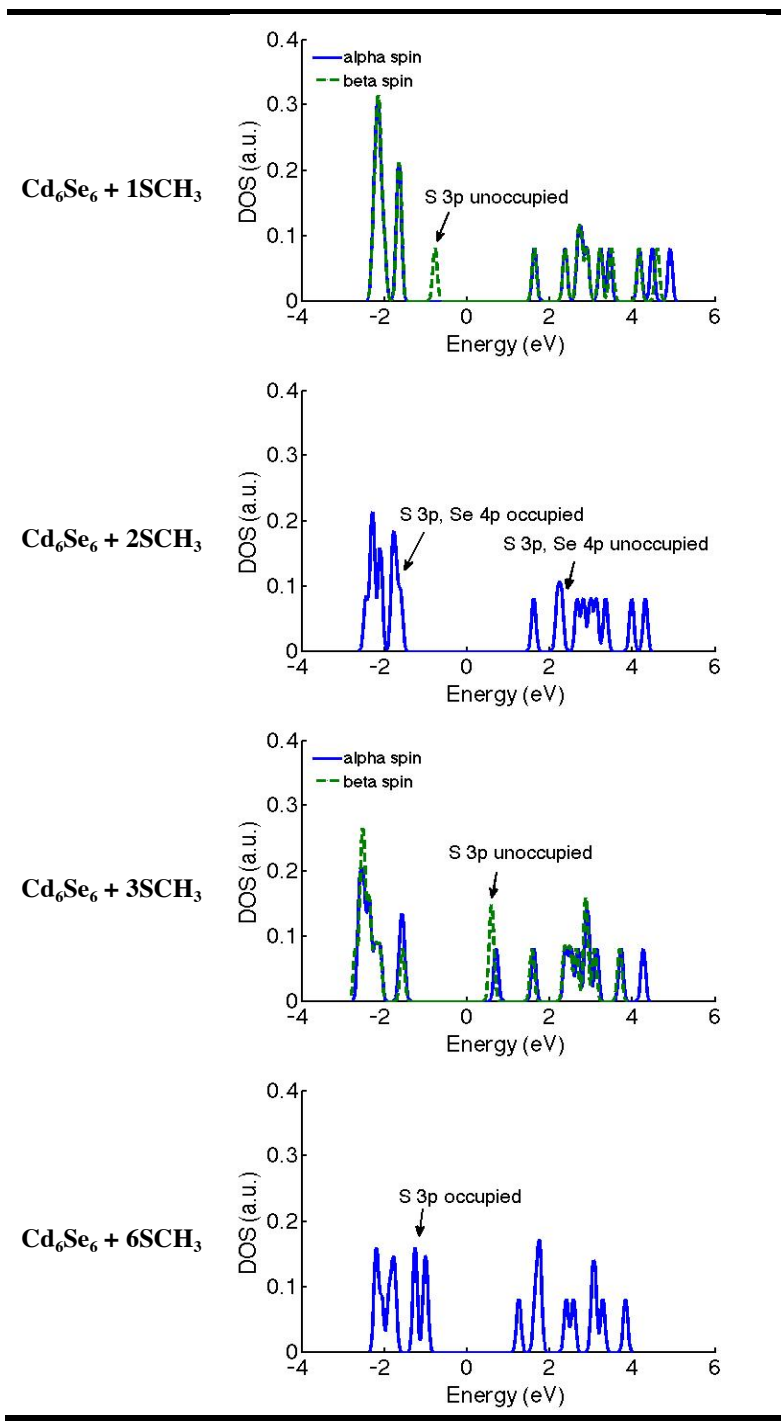

Table 7. Initial and Optimized structures of $Cd_{13}Se_{13}$ + $1SCH_3$ using the B3LYP functional with the LANL2DZ/6-31G* (CdSe/Ligand) basis sets (Cd: cyan, Se: yellow, H: white, S: orange, C: gray).

|  | Position 1 | Position 2 | Position 3 |
| --- | --- | --- | --- |

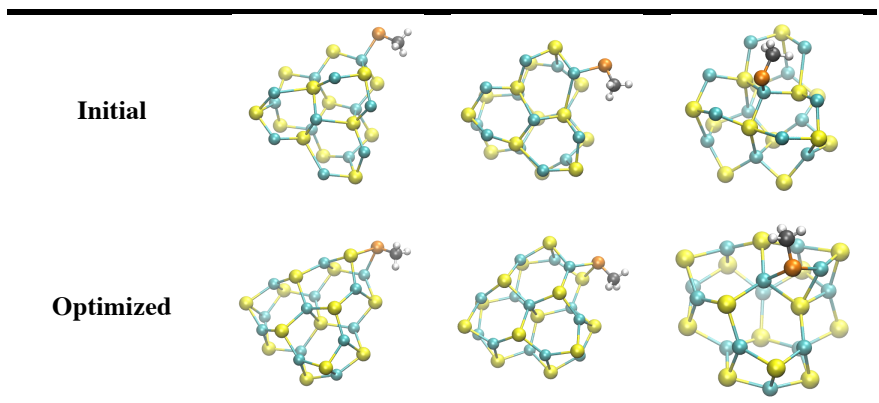

## 3.3. MPA, Cys and the Shorter Chain Analogues

As discussed in Section 2.1, MPA and Cys both contain thiol and carboxylic acid terminal group. We noted that when thiol was deprotonated into thiolate and attached to the CdSe surface, the carboxylic group was also deprotonated.[9] The -COO$^-$ group was usually attached to the surface of TiO$_2$,[7,9,14] though experimental work has also found that some portion of the -COO$^-$ group could also be attached to the CdSe surface.[9] In our simulation, we assumed the -COO$^-$ group was mainly attached to the surface of TiO$_2$. However, instead of involving the TiO$_2$ QD into our calculation, we just simplified the model by keeping the carboxyl group in the form of -COOH. The amine group, which was in the form of -NH$_2$, could also interact with the CdSe surface. The effects of these various functional groups on the QD-ligand systems are discussed below.

The Cd$_6$Se$_6$ was fully saturated with the thiol-radicals of MPA, Cys and their shorter chain analogue and each Cd was bonded with an organic ligand (Table 8). Then the composite systems were relaxed to their lowest energy configuration. As compared with MPA and Cys, reducing the length of the alkane chain resulted in the formation of a 15% smaller energy gap and a 6% shorter Cd-S bond (Table 9). As we can see from Table 8, the relaxed systems were all presented as open structures. The $Cd_6Se_6$-Cys system possessed two Cd-S-C-C-N-Cd rings and no O-Cd bond has formed either for the $Cd_6Se_6$-Cys system or for the $Cd_6Se_6$-MPA complex. For the other shorter chain ligands, the CdSe–SCH($NH_2$)COOH system is associated with two Cd-S-C-C-N-Cd rings and one Cd-S-C-C-O-Cd ring, whereas the CdSe-$SCH_2$COOH system has created one Cd-S-C-C-O-Cd ring. The coordinates of the optimized systems were documented in an independent file named as coordinates.txt (Supplementary Material). As we can see, when Cd-N and Cd-O bonds were generated, we observed a stronger binding between the thiol-radical ligands and the CdSe surface atoms (Table 9).

After analyzing the relaxed structures and BEs of the composite systems, we probed the density of states (DOS) and excitation energies of the complexes. As we can see from the second column of Table 10, the frontier orbitals were mainly composed by the ligand orbitals. These ligand states originated a series of small absorption peaks at a lower energy range (Table 10, column two).

In this subsection, we found that both amine and carboxyl show a great affinity to bond with the neighboring Cd atom, especially when we reduced the length of the ligand. (Table 8 and Table 9).

Table 8. Optimized structures of $Cd_6Se_6$ + 6ligands using the B3LYP functional with the LANL2DZ/6-31G* (CdSe/Ligand) basis sets (Cd: cyan, Se: yellow, H: white, S: orange, C: gray, O: red, N: blue).

| $SCH_3$ | Cys | MPA | SCH($NH_2$)COOH | $SCH_2$COOH |
|---|---|---|---|---|

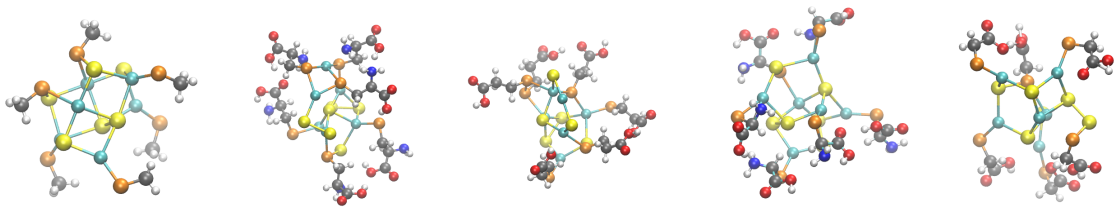

Table 9. BE (kcal/mol), bond length (d(Cd-S) and d(Se-S)) (Å) and HOMO-LUMO Gap (eV) of optimized $Cd_6Se_6$ fully saturated by 5 type thiol-radical ligands, using the B3LYP functional with the LANL2DZ/6-31G* (CdSe/Ligand) basis sets.

|  | $Cd_6Se_6$ | $SCH_3$ | Cys | MPA | $SCH(NH_2)COOH$ | $SCH_2COOH$ |
|---|---|---|---|---|---|---|
| **BE** | -- | -25.21 | -28.87 | -26.46 | -31.99 | -32.38 |
| *d(Cd-S)* | -- | 2.456 | 2.637 | 2.670 | 2.510 | 2.493 |
| *d(Se-S)* | -- | 2.346 | 2.345 | 2.334 | 2.327 | 2.322 |
| **HOMO-LUMO Gap** | 3.14 | 2.23 | 2.76 | 2.5 | 2.33 | 2.06 |

Table 10. DOS and TDDFT spectra plots of optimized bare $Cd_6Se_6$ and fully saturated $Cd_6Se_6$ with 5 types thiol-radical ligands. The B3LYP / (LANL2DZ/6-31G*) method is used for the TDDFT calculation. A Gaussian broadening of 0.05 eV has been utilized. The black arrows presented the localization of ligand orbitals.

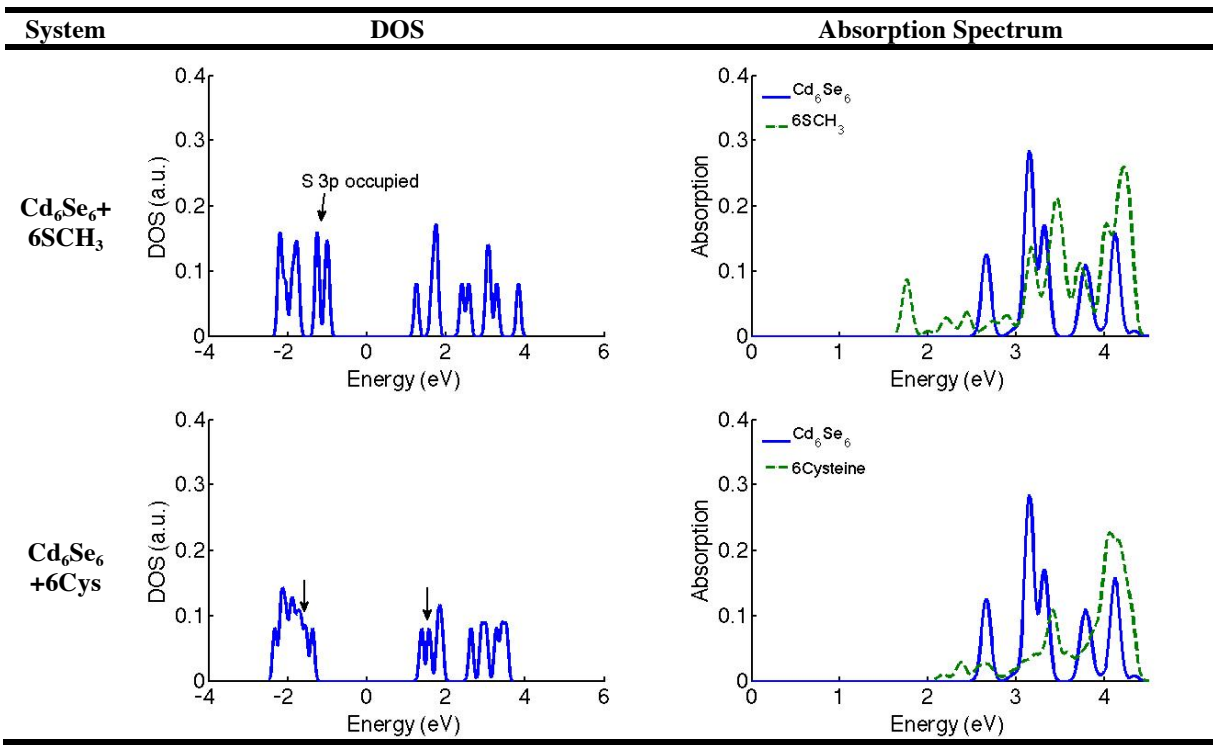

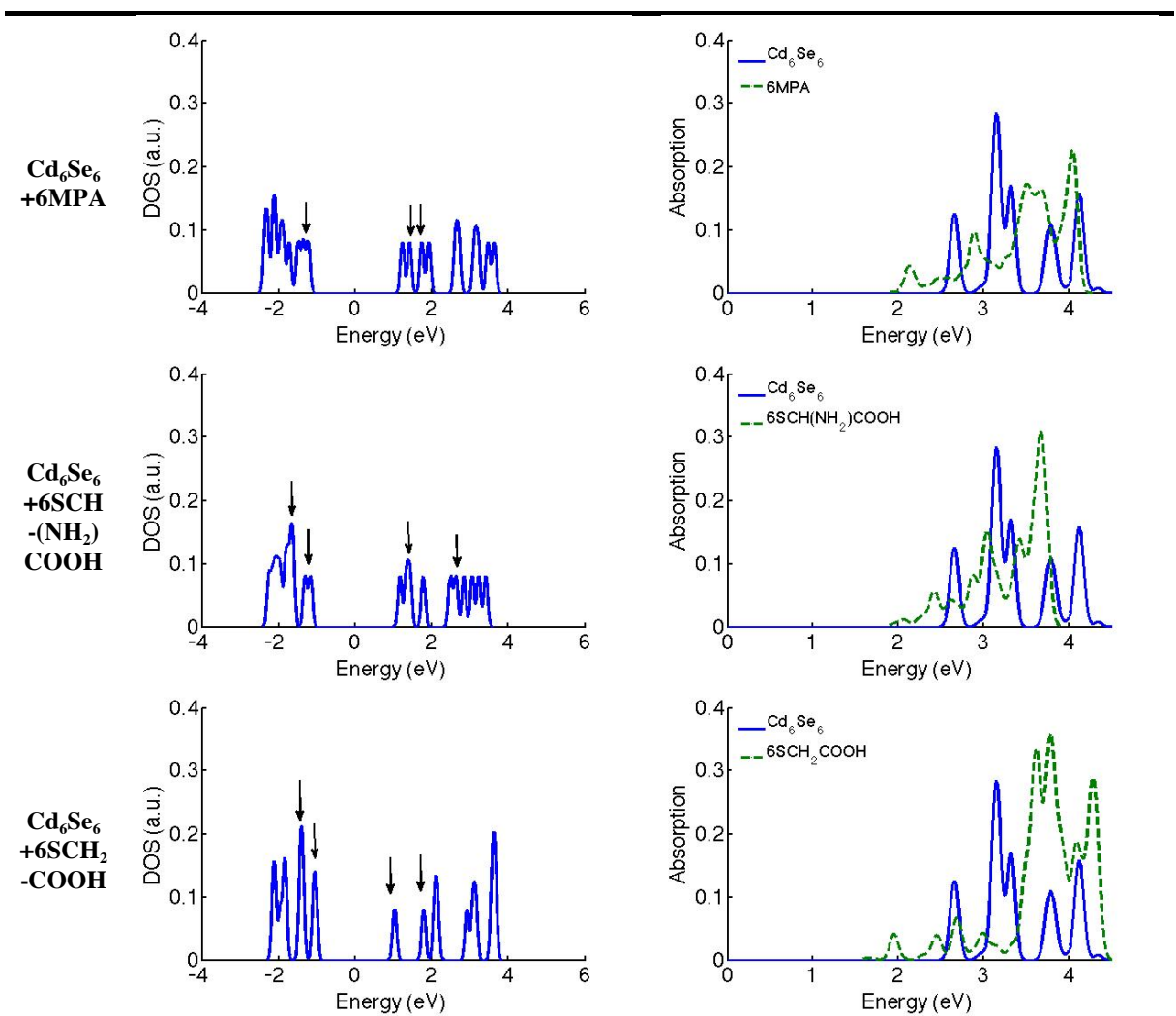

## 4. Conclusions

In this work, we performed a first-principles study of small $Cd_nSe_n$ QDs ('n' = 6, 13, and 33) with MT, Cys and MPA and their reduced-chain analogues. The major conclusions of the paper were as follows:

1) Thiol-radicals ligands were more influential in determining the structure and electronic properties of the CdSe QDs as compared with thiol ligands. The sulfur 3p orbitals were localized in the frontier orbitals (Table 2), which altered the absorption spectrum of bare $Cd_6Se_6$ by inducing a new lower energy absorption peak (Figure 2);

2) The thiol-radical-ligated complex was very sensitive to the position and number of ligands (Table 4).

3) Both amine group on Cys and carboxyl group on MPA and Cys evinced a strong tendency to bond with the neighboring Cd atom, especially when the length of ligand was reduced. The formation of Cd-N and Cd-O bonds resulted in smaller HOMO-LUMO gaps and a stronger binding between the ligands and the surface atoms of CdSe nanoclusters (Table 8 and Table 9).

**Acknowledgment**

This work was supported by a BNL-SBU seed grant. Research (including funds for ALT and SSW) was supported by the U.S. Department of Energy, Basic Energy Sciences, Materials Sciences and Engineering Division. This research used resources of the National Energy Research Scientific Computing Center, which was supported by the Office of Science of the U.S. Department of Energy under Contract No. DE-AC02-05CH11231. This research also utilized resources of the New York Center for Computational Sciences at Stony Brook University/Brookhaven National Laboratory which was supported by the U.S. Department of Energy under Contract No. DE-AC02-98CH10886 and by the State of New York. The authors thank Dr. James W. Davenport of the Materials Sciences and Engineering Division of the U.S. Department of Energy and Dr. Yan Li of the Computational Science Center at Brookhaven National Laboratory for their invaluable comments.